\documentclass[aps,pre,twocolumn]{revtex4}
\usepackage{graphicx}
\usepackage{amsmath}
\usepackage{amssymb}
\usepackage{amsthm}
\usepackage{array}
\usepackage[pdftitle={Shen},colorlinks=True,
urlcolor=Violet,filecolor=MidnightBlue]{hyperref}
\usepackage{xy}
\newcommand{\erf}{\mbox{erf}}
\newcommand{\eps}{\epsilon}

\newcommand{\p}{\partial}
\newcommand{\al}{\alpha}

\newcommand{\vecR}{\vec{R}}
\newcommand{\vecr}{\vec{r}}

\begin{document}
\title{Statistical Mechanics of a Cat's Cradle}
\date{\today}

\begin{abstract}
It is believed that, much like a cat's cradle, the 
cytoskeleton can be thought of
as a network of strings under tension. 
We show that both regular and random bond-disordered
networks having 
bonds that buckle upon compression exhibit a variety of 
phase transitions as a function of 
temperature and extension.
The results of self-consistent
phonon calculations 
for the regular 
networks agree very well with computer simulations at
finite temperature.  The analytic theory also yields a
rigidity onset (mechanical percolation) and the fraction of extended bonds 
for random networks. There
is very good agreement with
the simulations by Delaney {\em et al.} 
(Europhys. Lett. 2005). 
The mean field theory reveals a 
nontranslationally invariant phase with 
self-generated heterogeneity of tautness, representing
``antiferroelasticity''.

\end{abstract}
\author{Tongye Shen and  Peter G. Wolynes}
\affiliation{Department of Chemistry \& Biochemistry,
Department of Physics, and Center for Theoretical 
Biological Physics,
University of California, San Diego,
La Jolla, California, 92093-0371}


\maketitle
It is thought that the cytoskeleton consists 
of a polymeric network which is 
maintained under tension~\cite{Ingber06,CCI98,WNSFM01}, much like the 
child's game made of strings known as a ``cat's cradle''. 
Unlike a macroscopic ``cat's cradle'', the 
cytoskeleton is subject to fluctuating forces  
from both thermal motions and the nonequilibrium noise due to 
motor proteins and constant polymerization and 
depolymerization events~\cite{pollard}.

Upon freezing molecular fluids that resist 
compression undergo transitions to
rigid solids. Here we show that, 
a caricature of the cytoskeleton, a model of 
interconnecting ropes that easily buckle also
undergoes a variety of phase transitions.
We study networks of ropes that bear tension when they
are stretched beyond a critical threshold, but otherwise buckle.
While the model's interactions contrast
 with the usual compression-bearing
mechanical interaction between atoms or
molecules, the ``inverted''
nonlinearity of ropes still leads to an onset of
rigidity like that in crystals or glasses, but when
the system is expanded rather than compressed.
Depending on where we set the baseline
of environmental pressure or tension, 
we may also say the system 
buckles upon compression. 

The cytoskeleton actually has both 
tension-bearing and compressional elements. The present theory,
can be extended to include both elements. 
Elsewhere we have discussed
the nonequilibrium aspects~\cite{SW05} that are also needed to capture
the essential features of the tensegrity 
model~\cite{WNSFM01,CCI98} of cell 
mechanics~\cite{FMBGN01}, but here we limit ourselves
to the effects of equilibrium fluctuations.

Delaney, Weaire, and Hutzler pioneered the study of
such buckling lattices with neighboring sites that interact with 
an ``inside-out'' potential  
$u(r;l)= \Theta (r-l) \cdot {k\over 2} \cdot (r-l)^2 $, 
where $\Theta()$ is the Heaviside unit step function~\cite{DWH05}.
The energy stored in a bond vanishes 
when the length of the bond $r$ is less than a critical value 
$l$, but increases quadratically beyond the cutoff $l$.
A two (three) dimensional network of
bonds is thermodynamically characterized by its
tension (negative pressure),  
as a function of its area (volume)
$-p_2= {\p F / \p s}$ ($-p_3={\p F/ \p v}$).

In the low temperature limit, 
the entropy term can be ignored and 
elementary statics suffices to describe
regular networks.
The square network, for example, has a tension
as a function of the area expansion given by
$-p_2^{sq}(s/s_0)= k(1- \sqrt{s_0/s})~\Theta(s/s_0 -1)$.
Rigidity only occurs for areas exceeding $s_0$.
Tensions for other lattices can also be computed
$p_2^{tri}=2\sqrt{3} p_2^{sq}$ and 
$p_2^{hex}=p_2^{tri}/3$ for the triangular and the hexagonal 
network respectively. 
For the 3D cubic network one finds  at $T=0$,
$-p_3^{sc}(v/v_0)= 2 k(1- \sqrt[3]{v_0/v}) \sqrt[3]{v_0/v}~\Theta(v/v_0 -1)$
yielding a rigidity threshold at $v_0$.
We have set $s_0$ (or $v_0$) as the area (or the volume) of the 
corresponding network constructed from all bonds at the unit length.   
The tension of 3D networks has a maximum 
as a function of volume, 
because at large volume the energy 
density of the system decreases as $v^{-1/3}$. 

At high temperature or equivalently for soft springs,
even without disorder, static models are insufficient and  
one always needs 
a {\em statistical} mechanical description. 
If the network is disordered, i.e., having
a broad distribution of onset length $l$, 
finding the tension is a nontrivial statistical problem
even in the limit of low temperature~\cite{DWH05}.  

To analytically treat statistical cat's cradles,
we adopt the self-consistent 
phonon technique, initially developed for
anharmonic lattice solids with compression bearing 
interactions~\cite{Fixman69,Cho67}.
In this approach one replaces the pairwise interaction in
the equilibrium density matrix   
$\exp[-\beta \sum\sum_{ij} u_{ij}(\vecr_i-\vecr_j)]$ 
by a nearly equivalent {\em effective single-body}
term as $\Pi_i \exp[- \tilde {u}_i(\vecr_i-\vecR_i)]$. 
Decoupling the manybody effects can be carried out by assuming
a self-consistent one-body potential $\tilde{u}$ has a quadratic form
$\al (\vecr-\vecR)^2$. 

The self-consistent phonon method has not only been applied
to crystals but also to 
noncrystalline solids~\cite{SW84}, network glasses~\cite{HW03},
and to problems of nonequilibrium structural assembly~\cite{SW04}.
The self-consistent phonon method can be made the 
beginning of a systematic expansion~\cite{Fixman69}, 
but already has been shown to work very 
well for systems made up of compressive elements.

We study both 2D and 3D systems. Long wavelength phonons lead
to a diverging mean square
displacement (MSD) with increasing system size for 2D 
systems, in a strict sense.
However this divergence is weak (logarithmic) 
and there is still a crossover reflecting 
rather rapid decrease in fluctuations of
the mean square nearest neighbor distance upon extension.
In fact, the self-consistent phonon method,
when used for the 2D system 
predicts the correct
onset of rigidity  for the random 2D network
at $T=0$ as we shall see.

In the self-consistent phonon theory the 
partition function 
$z(\xi) =[ \int d\vecr \exp(-\eps(\xi)  - \al(\xi) r^2)]^N$
is evaluated as a function of expansion 
ratio $\xi= s/s_0$ or $v/v_0$. 
Both the on-site free energy $\eps$ and the phonon frequency $\al$ 
depend on the ratio of the mean neighbor distance $R$ to the critical
extension $l$, $R/l= \sqrt[D]\xi$. 
We can use the chain rule to calculate tension, e.g.,
 $-p_2={\p f\over \p s}=-{\p\ln z(\xi(R))\over \beta\p s(R)}$.

In three dimensions the Mayer $f$-bond $v(\vecR; l,\al)$ 
for the interaction $u(r;l)$ when averaged over
the Gaussian fluctuation of a neighboring particle 
around its mean position $\vecR$ having its own spring constant $\al$
gives
$$\exp[-\beta v(R)]\equiv ({\pi\over \al})^{-{3\over 2}}\int r^2 dr \int d\Omega 
e^{-\al (\vecr- \vecR)^2 - \beta u(r)}$$
This can be expressed using the error function $\erf()$.
The 2D averaged Mayer $f$-bond can 
be expressed as a numerical integral of 
modified Bessel function $I_0$.

An interesting property that can
be calculated within the self-consistent phonon theory 
is the fraction of the bonds bearing force (that are taut),
i.e., having a length larger than
the intrinsic cutoff $l$.
These are functions of $\al$ and $R$:
$q_2 = 2a  e^{-\al R^2} \int_l^\infty r dr I_0(2 \al R r ) e^ {-\al r^2}$
and $q_3= 1+ {1\over 2}\erf[(R-l)\sqrt{\al}] 
- {1\over 2}\erf[(R+l)\sqrt{\al}]
+{(\al\pi)^{-1/2}\over 2R} [e^{-\al (R-l)^2} - e^{-\al (R+l)^2}]$
for 2D and 3D cases.

\begin{figure}%
\includegraphics[scale=0.70, angle=0]{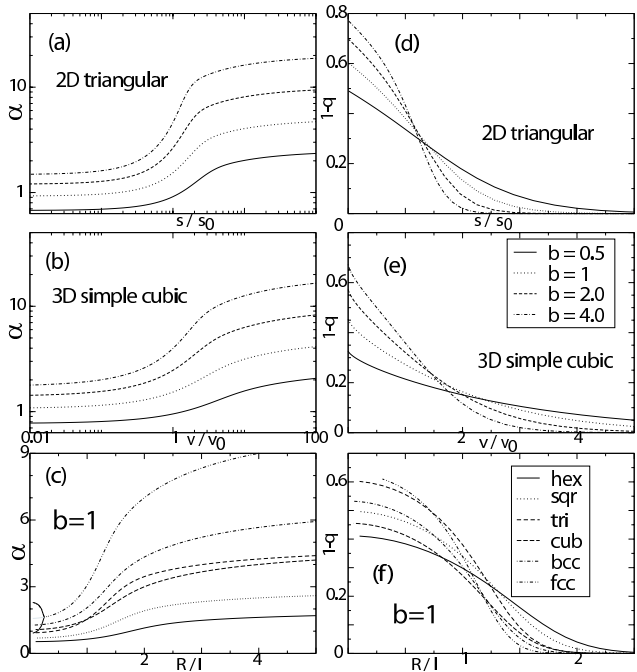}
\caption{\label{svaq} The phonon frequencies $\al$
are shown in the left panels as functions of 
the area (a) or volume (b) of networks for various $b=\beta k/2$,
and as functions of the
linear expansion parameter with constant $b=1$ for various types
of networks (c).
The fractions of free bonds $1-q$ 
of corresponding setups are shown in the right panels (d)-(f). 
Note that (a), (b), (d), and (e) share a legend 
while (c) and (f) share the other legend.}
\end{figure}

The self-consistent phonon frequency 
$\al$ for several different regular 
networks is shown in the left panels of Fig.~\ref{svaq}.  
Unlike the situation for the systems made of hard-spheres~\cite{SW05}, 
there is always some nonzero self-consistent solution $\al$.
Entropy always leads to some rigidity. There is, however,
a rapid crossover from low
$\al$ to high $\al$ 
accompanying  the expansion of the system.
Generally either high coordination
number $z$ or larger microscopic rigidity-temperature
ratio $b$ ($=k \beta / 2$) lead to a larger $\al$.
The fraction of ``loose'' bonds,
$(1-q)$ predicted from the phonon theory 
is plotted in the right panels of Fig.~\ref{svaq}.
This fraction decreases as the network expands.

\begin{figure}%
\includegraphics[scale=0.6, angle=0]{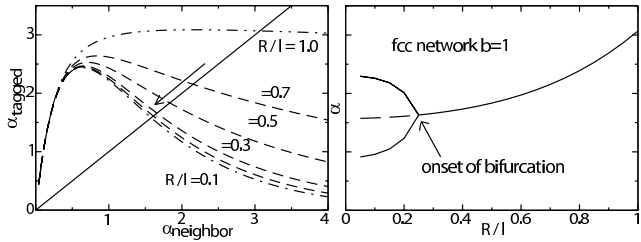}
\caption{\label{scp}
The illustration of the onset
of bifurcation solutions for a specific case.}
\end{figure}

We have shown there is a 
distinct bifurcation of $\al$ that also occurs 
when the volume ratio $\xi$ becomes small enough for 
3D networks with high rigidity-temperature ratio 
$b$, This  bifurcation appears
in all three types of 3D networks studied here.
The fcc networks most clearly display
this bifurcation, followed by the bcc, and then the sc
networks at smaller $\xi$ and higher $b$. 
As shown in Fig.~\ref{scp},
usually there is only one stable solution of
the self-consistent equation
$\al_{tagged}= f(\al_{neighbor})$. 
However, under certain
conditions (small volume and large rigidity-temperature ratio, or
pictorially, when the slope at the intercept is not in the range of
between -1 and 1), the self-consistent solution becomes unstable,
and another {\em pair} of solutions $\al_l$ and $\al_h$
satisfying $\al_l= f(\al_h)$ and $\al_h=f(\al_l)$ emerges
as the stable solution. This
stable pair solution
indicates the possibility of having a nontranslationally invariant solution,
in which a low $\al$ particle
is surrounded by high $\al$ neighbors and
vice versa. The resulting spatial heterogeneity is self-generated.
In the bcc and sc networks, this solution can be easily interpreted 
physically: there is a symmetry-broken 
phase of bcc or sc networks
that can be thought as an ``antiferromagnetic'' (AF) one
with the underlying lattice being split to two subnetworks
(sc $\to$ fcc or bcc $\to$ sc):
one of the two equal subnetworks bears more tension than the 
other. We call this the ``antiferroelastic'' phase. In the fcc case,
since there are four particles in the unit
cell, one high $\al$ particle cannot
be completely surrounded by low $\al$ particles.
Nevertheless one can have two possible ordered AF ground states 
similar to the crystal structures of AuCu and AuCu$_3$. 
To test this possibility, we set up a more elaborate 
self-consistent solution having four values of 
$\al =(\al_0, \al_x, \al_y, \al_z)$
at (000), (011), (101), (110) respectively,
$ \al_0 = f(\al_x,\al_y,\al_z),$
$ \al_x = f(\al_0,\al_z,\al_y),$
$ \al_y = f(\al_z,\al_0,\al_x),$
$ \al_z = f(\al_y,\al_x,\al_0).$
By explicitly
labeling all four particles in a unit cell
(as done similarly in the AF Ising fcc
system~\cite{BR05}), we searched but failed to find these 
explicit symmetry-broken solutions. 
Thus we believe this frustration of fcc network
would lead to a random phase which is analogous to a self-generated 
glass. This possibility may potentially be resolved using a 
replica calculation~\cite{M95,MP99}.

\begin{figure}%
\includegraphics[scale=0.32, angle=0]{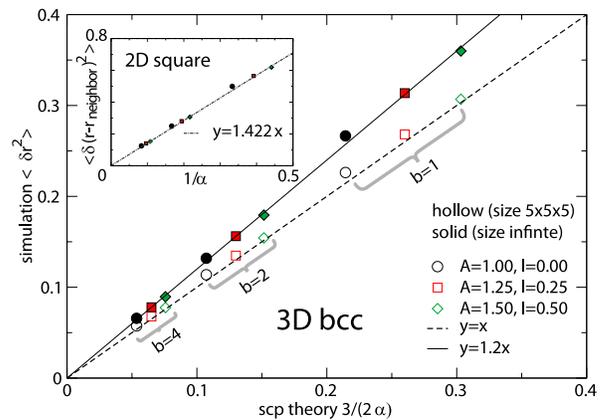}
\caption{\label{cpts} The comparison of the
MSD of bcc networks between theory and  
simulation for various parameters $R\equiv\sqrt{3}/2\times A$, 
$b$, and $l$. The corresponding results for square networks are shown
in the inset ($R\equiv A$).}
\end{figure}

We checked our self-consistent phonon theory using 
explicit computer simulations of the network with thermal
noise. We show the results for the bcc case  in
Fig.~\ref{cpts}. The simulations were performed with
stochastic dynamics on $n^3$ unit cells with periodic boundary 
condition for various rigidity-temperature 
ratios $b$ and nonlinearity parameters $l$ for various values of $n$.
We see the self-consistent phonon theory agrees quite well with the simulation.
For the simulation of size $5^3$,
the errors for the MSD of the particle position
between theory and simulation
are only about 1\%.
We extrapolated to the thermodynamic limit based on
a series of simulation of different sizes of networks.
A Similar good agreement is achieved for 2D networks
when the MSD of {\em relative neighbor} position from simulation is
compared with theory.

To check the prediction of spontaneous
symmetry breaking of neighboring $\al$ values
requires a large ratio of $l/R\sim 10$
necessitating simulation using a very large system
at very low temperature. We estimate the necessary 
system size to be at least $10^6$. We were unable to 
carry out such a large scale simulation.

\begin{figure*}%
\includegraphics[scale=1.0, angle=0]{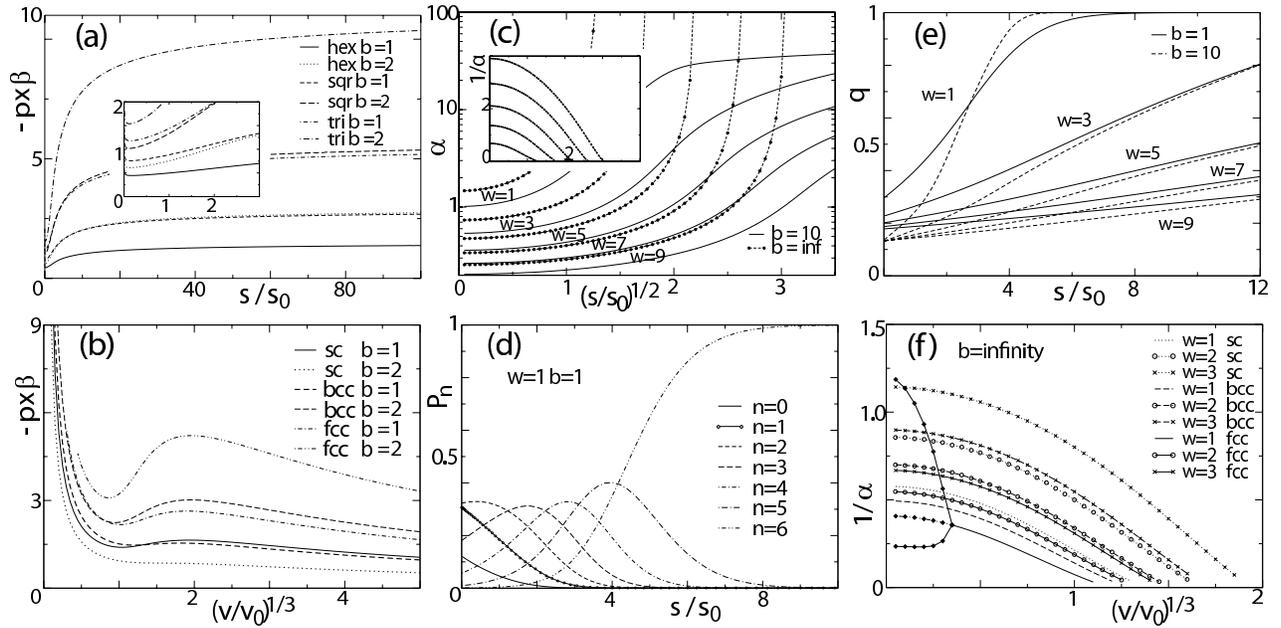}
\caption{\label{pressure} The applied tensions
as functions
of area expansion of 2D triangular 
network for various $b$ are shown in (a) with
the zoom-in at the small $s/s_0$ region shown as an inset.
We show the tensions of 3D networks in (b).
The random bond 2D triangular network results are
shown in (c)-(e). The effects of the random bond distribution 
$w$ on the phonon frequency
$\al$ are shown in (c) with the inversion $1/\al$ shown
in the inset of (c) to better illustrate the critical
area for onset of rigidity.
The portion of vertices with $n$ bonds tauted $P_n$ 
and the mean taut ratio $q$ are shown in (d) and (e) respectively. 
We show the inversion $1/\al$ as functions of volume 
for various 3D random networks in (f). }
\end{figure*}

We plot the tension vs.~extension in Fig.~\ref{pressure} (a) and (b). 
Above a critical tension, there are two solutions
for $p(s/s_0)=p$.
The solution
with ${\p p/ \p s}<0$ is
thermodynamically stable for a constant
pressure ensemble. 
The system always appears to be stable
in two dimensions.
For 3D networks,
As shown in Fig~\ref{pressure}(b), there is single (unstable) solution
at high temperature for constant $p_3$; but there are three
(one stable and two unstable)
solutions at low temperature.
The values of the critical rigidity-temperature ratio 
for this transition 
are $b^{*}=1.04$, 0.76, and
0.48 for sc, bcc, and fcc
networks respectively. Thus the
network is intrinsically unstable for the constant $p$ ensemble
for high temperature or
low spring constant, i.e. $b< b^*$,
regardless of supplied tension $-p$.
For $b> b^*$, there is a critical
tension $-p^*$ above which the system is 
no longer stable. Generally a larger $b$ 
implies  a larger stable region.

In the large $s/s_0$ or $v/v_0$ limit,
the Dulong-Petit law 
provides pressure that can be 
based on dimension 
counting alone, giving $p_2-p^{o}=N k_B T /(2 S)$
and $p_3- p^{o}=N k_B T /(2 V)$.
Here $p^{o}=p(T=0)$ is the pressure from the mechanical static calculations
at zero temperature.
The factor ${1\over 2}$ arises
because only the potential energy is counted here, the kinetic energy
counts the other half. Thus for large expansion ratios, $p_2\to \mbox{const.}$
and $p_3\to 0$, which are consistent with the tension curves 
shown in Fig.~\ref{pressure}(a) and (b).

So far we have focussed on the effects of 
thermal disorder alone. We now discuss the effects of
bond disorder. For illustration, we concentrate on 
the 2D random network,
especially in the $\beta k\to \infty$ limit. 
Following Delaney {\em et al.}~\cite{DWH05}, we 
studied a ``flat'' random distribution
of string lengths. 
More general distributions
present no new challenges to our method.

As shown in Fig.~\ref{pressure}(c) for the triangular network,
phonon frequencies $\al$ rise quickly with
expansion at zero temperature when $b=\infty$.
The smallest area $s^*/s_0$ satisfying 
$\al^{-1}(s/s_0)=0$ is the critical value
for the onset of macroscopic rigidity (mechanical percolation).
Since the spring 
constant $k$ and temperature are always bundled together 
in the treatment, once the system
is taut, $\al \to \infty$ for $1/b=0$. 
For finite temperature, $\al$ is always finite.
Th rigidity onset $s^*/s_0$ calculated here
quantitatively agrees with the simulation results
in Ref.~\cite{DWH05} as a function of 
width of distribution of string length $w$. 
As expected, a larger $w$ means a
lower $\al$ generally.
We also display in Fig.~\ref{pressure}
the fraction of vertices with
given number of taut bonds $P_n$ (d)
and the mean fraction of taut bonds $q$ (e).
These properties have also been 
obtained by the simulations of Delaney {\em et al.}
We can only make a qualitative comparison for these two properties
since our calculations were performed with finite
$b$, corresponding to $T\ne 0$ 
while the simulations were strictly done at $T=0$.
Our theory does not allow us to calculate the loose bond fraction
at $T=0$ due to the bundling of $\beta$ and $k$.
An approximation was made
for the calculation of $p(n)$, the fraction of vertices with
given number $n$ bonds taut  based on the mean fraction
of taut bonds $q$, i.e.,
$p(n)={Z!\over n! (Z-n)!} q^n (1-q)^{Z-n}$. Here $Z$ is
the coordination number.

As shown in Fig.~\ref{pressure}(f),  
for bond-disordered 3D networks, again there is a 
rigidity onset, which is similar to those of 2D networks.
Among the cases studied only
the $w=1$ fcc network showed the antiferroelastic
structural inhomogeneity bifurcation. 
The self-consistent phonon method suggests
random bond-disorder
destroys antiferroelasticity.
This work was supported in part by NSF-CTBP.

\bibliographystyle{apsrev}

\bibliography{cats,cyto}

\begin{thebibliography}{15}
\expandafter\ifx\csname natexlab\endcsname\relax\def\natexlab#1{#1}\fi
\expandafter\ifx\csname bibnamefont\endcsname\relax
  \def\bibnamefont#1{#1}\fi
\expandafter\ifx\csname bibfnamefont\endcsname\relax
  \def\bibfnamefont#1{#1}\fi
\expandafter\ifx\csname citenamefont\endcsname\relax
  \def\citenamefont#1{#1}\fi
\expandafter\ifx\csname url\endcsname\relax
  \def\url#1{\texttt{#1}}\fi
\expandafter\ifx\csname urlprefix\endcsname\relax\def\urlprefix{URL }\fi
\providecommand{\bibinfo}[2]{#2}
\providecommand{\eprint}[2][]{\url{#2}}

\bibitem[{\citenamefont{Ingber}(2006)}]{Ingber06}
\bibinfo{author}{\bibfnamefont{D.~E.} \bibnamefont{Ingber}},
  \bibinfo{journal}{FASEB J.} \textbf{\bibinfo{volume}{20}},
  \bibinfo{pages}{811} (\bibinfo{year}{2006}).

\bibitem[{\citenamefont{Chicurel et~al.}(1998)\citenamefont{Chicurel, Chen, and
  Ingber}}]{CCI98}
\bibinfo{author}{\bibfnamefont{M.~E.} \bibnamefont{Chicurel}},
  \bibinfo{author}{\bibfnamefont{C.~S.} \bibnamefont{Chen}}, \bibnamefont{and}
  \bibinfo{author}{\bibfnamefont{D.~E.} \bibnamefont{Ingber}},
  \bibinfo{journal}{Curr Opin. Cell Biol.} \textbf{\bibinfo{volume}{10}},
  \bibinfo{pages}{232} (\bibinfo{year}{1998}).

\bibitem[{\citenamefont{Wang et~al.}(2001)\citenamefont{Wang, Naruse,
  Stamenovic, Fredberg, Mijailovich, Tolic-Norrelykke, Polte, Mannix, and
  Ingber}}]{WNSFM01}
\bibinfo{author}{\bibfnamefont{N.}~\bibnamefont{Wang}},
  \bibinfo{author}{\bibfnamefont{K.}~\bibnamefont{Naruse}},
  \bibinfo{author}{\bibfnamefont{D.}~\bibnamefont{Stamenovic}},
  \bibinfo{author}{\bibfnamefont{J.~J.} \bibnamefont{Fredberg}},
  \bibinfo{author}{\bibfnamefont{S.~M.} \bibnamefont{Mijailovich}},
  \bibinfo{author}{\bibfnamefont{I.~M.} \bibnamefont{Tolic-Norrelykke}},
  \bibinfo{author}{\bibfnamefont{T.}~\bibnamefont{Polte}},
  \bibinfo{author}{\bibfnamefont{R.}~\bibnamefont{Mannix}}, \bibnamefont{and}
  \bibinfo{author}{\bibfnamefont{D.~E.} \bibnamefont{Ingber}},
  \bibinfo{journal}{Proc. Natl. Acad. Sci. USA.} \textbf{\bibinfo{volume}{98}},
  \bibinfo{pages}{7765} (\bibinfo{year}{2001}).

\bibitem[{\citenamefont{Pollard and Earnshaw}(2002)}]{pollard}
\bibinfo{author}{\bibfnamefont{T.~D.} \bibnamefont{Pollard}} \bibnamefont{and}
  \bibinfo{author}{\bibfnamefont{W.~C.} \bibnamefont{Earnshaw}},
  \emph{\bibinfo{title}{Cell biology}} (\bibinfo{publisher}{W.B. Saunders},
  \bibinfo{address}{New York}, \bibinfo{year}{2002}).

\bibitem[{\citenamefont{Shen and Wolynes}(2005)}]{SW05}
\bibinfo{author}{\bibfnamefont{T.}~\bibnamefont{Shen}} \bibnamefont{and}
  \bibinfo{author}{\bibfnamefont{P.~G.} \bibnamefont{Wolynes}},
  \bibinfo{journal}{Phys. Rev. E} \textbf{\bibinfo{volume}{72}},
  \bibinfo{pages}{041927} (\bibinfo{year}{2005}).

\bibitem[{\citenamefont{Fabry et~al.}(2001)\citenamefont{Fabry, Maksym, Butler,
  Glogauer, Navajas, and Fredberg}}]{FMBGN01}
\bibinfo{author}{\bibfnamefont{B.}~\bibnamefont{Fabry}},
  \bibinfo{author}{\bibfnamefont{G.~N.} \bibnamefont{Maksym}},
  \bibinfo{author}{\bibfnamefont{J.~P.} \bibnamefont{Butler}},
  \bibinfo{author}{\bibfnamefont{M.}~\bibnamefont{Glogauer}},
  \bibinfo{author}{\bibfnamefont{D.}~\bibnamefont{Navajas}}, \bibnamefont{and}
  \bibinfo{author}{\bibfnamefont{J.~J.} \bibnamefont{Fredberg}},
  \bibinfo{journal}{Phys. Rev. Lett.} \textbf{\bibinfo{volume}{87}},
  \bibinfo{pages}{148102} (\bibinfo{year}{2001}).

\bibitem[{\citenamefont{Delaney et~al.}(2005)\citenamefont{Delaney, Weaire, and
  Hutzler}}]{DWH05}
\bibinfo{author}{\bibfnamefont{G.~W.} \bibnamefont{Delaney}},
  \bibinfo{author}{\bibfnamefont{D.}~\bibnamefont{Weaire}}, \bibnamefont{and}
  \bibinfo{author}{\bibfnamefont{S.}~\bibnamefont{Hutzler}},
  \bibinfo{journal}{Europhys. Lett.} \textbf{\bibinfo{volume}{72}},
  \bibinfo{pages}{990} (\bibinfo{year}{2005}).

\bibitem[{\citenamefont{Fixman}(1969)}]{Fixman69}
\bibinfo{author}{\bibfnamefont{M.}~\bibnamefont{Fixman}}, \bibinfo{journal}{J.
  Chem. Phys.} \textbf{\bibinfo{volume}{51}}, \bibinfo{pages}{3270}
  (\bibinfo{year}{1969}).

\bibitem[{\citenamefont{Choquard}(1967)}]{Cho67}
\bibinfo{author}{\bibfnamefont{P.~F.} \bibnamefont{Choquard}},
  \emph{\bibinfo{title}{The anharmonic crystal}}, Frontiers in physics
  (\bibinfo{publisher}{W.A. Benjamin, Inc.}, \bibinfo{address}{New York},
  \bibinfo{year}{1967}).

\bibitem[{\citenamefont{Stoessel and Wolynes}(1984)}]{SW84}
\bibinfo{author}{\bibfnamefont{J.}~\bibnamefont{Stoessel}} \bibnamefont{and}
  \bibinfo{author}{\bibfnamefont{P.~G.} \bibnamefont{Wolynes}},
  \bibinfo{journal}{J. Chem. Phys.} \textbf{\bibinfo{volume}{80}},
  \bibinfo{pages}{4502} (\bibinfo{year}{1984}).

\bibitem[{\citenamefont{Hall and Wolynes}(2003)}]{HW03}
\bibinfo{author}{\bibfnamefont{R.~W.} \bibnamefont{Hall}} \bibnamefont{and}
  \bibinfo{author}{\bibfnamefont{P.~G.} \bibnamefont{Wolynes}},
  \bibinfo{journal}{Phys. Rev. Lett.} \textbf{\bibinfo{volume}{90}},
  \bibinfo{pages}{085505} (\bibinfo{year}{2003}).

\bibitem[{\citenamefont{Shen and Wolynes}(2004)}]{SW04}
\bibinfo{author}{\bibfnamefont{T.}~\bibnamefont{Shen}} \bibnamefont{and}
  \bibinfo{author}{\bibfnamefont{P.~G.} \bibnamefont{Wolynes}},
  \bibinfo{journal}{Proc. Natl. Acad. Sci. USA} \textbf{\bibinfo{volume}{101}},
  \bibinfo{pages}{8547} (\bibinfo{year}{2004}).

\bibitem[{\citenamefont{Beath and Ryan}(2005)}]{BR05}
\bibinfo{author}{\bibfnamefont{A.~D.} \bibnamefont{Beath}} \bibnamefont{and}
  \bibinfo{author}{\bibfnamefont{D.~H.} \bibnamefont{Ryan}},
  \bibinfo{journal}{Phys. Rev. B} \textbf{\bibinfo{volume}{72}},
  \bibinfo{pages}{014455} (\bibinfo{year}{2005}).

\bibitem[{\citenamefont{Monasson}(1995)}]{M95}
\bibinfo{author}{\bibfnamefont{R.}~\bibnamefont{Monasson}},
  \bibinfo{journal}{Phys. Rev. Lett.} \textbf{\bibinfo{volume}{75}},
  \bibinfo{pages}{2847} (\bibinfo{year}{1995}).

\bibitem[{\citenamefont{Mezard and Parisi}(1999)}]{MP99}
\bibinfo{author}{\bibfnamefont{M.}~\bibnamefont{Mezard}} \bibnamefont{and}
  \bibinfo{author}{\bibfnamefont{G.}~\bibnamefont{Parisi}},
  \bibinfo{journal}{Phys. Rev. Lett.} \textbf{\bibinfo{volume}{82}},
  \bibinfo{pages}{747} (\bibinfo{year}{1999}).

\end{thebibliography}

\end{document}